\documentclass[onecolumn,floatfix,superscriptaddress,a4paper,
               showpacs,showkeys,nofootinbib,preprint]{revtex4}
\usepackage{epsfig}
\usepackage{latexsym}
\usepackage{xspace}
\usepackage{hyperref}
\usepackage[latin2]{inputenc}
\usepackage{indentfirst}
\usepackage{enumerate}

\usepackage{amsmath}
\usepackage{amssymb}
\usepackage[english]{babel}
\usepackage{url}

\newcommand{\eq}[1]{\begin{align} #1 \end{align}}

\newcommand{\bvar}{\mathbf}

\begin{document}

\title{Limiting temperature of pion gas \\with the van der Waals equation of state}
\author{R. V. Poberezhnyuk}
\affiliation{
Bogolyubov Institute for Theoretical Physics, 03680 Kiev, Ukraine}
\affiliation{
Frankfurt Institute for Advanced Studies, Johann Wolfgang Goethe University,
D-60438 Frankfurt, Germany}
\author{V. Vovchenko}
\affiliation{
Frankfurt Institute for Advanced Studies, Johann Wolfgang Goethe University,
D-60438 Frankfurt, Germany}
\affiliation{
GSI Helmholtzzentrum f\"ur Schwerionenforschung GmbH, D-64291 Darmstadt, Germany}
\affiliation{
Taras Shevchenko National University of Kiev, 03022 Kiev, Ukraine}

\author{D. V. Anchishkin}
\affiliation{
Bogolyubov Institute for Theoretical Physics, 03680 Kiev, Ukraine}
\affiliation{
Frankfurt Institute for Advanced Studies, Johann Wolfgang Goethe University,
D-60438 Frankfurt, Germany}
\affiliation{
Taras Shevchenko National University of Kiev, 03022 Kiev, Ukraine}

\author{M. I. Gorenstein}
\affiliation{
Bogolyubov Institute for Theoretical Physics, 03680 Kiev, Ukraine}
\affiliation{
Frankfurt Institute for Advanced Studies, Johann Wolfgang Goethe University,
D-60438 Frankfurt, Germany}

\date{\today}

\pacs{ 12.40.Ee, 12.40.-y}

\keywords{Pion gas, limiting temperature}

\begin{abstract}
The grand canonical ensemble formulation of the van
der Waals equation of state that includes the effects
of Bose statistics is applied to the equilibrium system of interacting pions.
If the attractive interaction between pions is large enough,
a limiting temperature $T_0$ emerges, i.e., no thermodynamical equilibrium
is possible at $T>T_0$. The system pressure $p$, particle number density $n$,
and energy density $\varepsilon$ remain finite at $T=T_0$, whereas
for $T$ near $T_0$
both the specific heat $C=d\varepsilon/dT$ and
the scaled variance of particle number fluctuations $\omega[N]$
are proportional to $(T_0-T)^{-1/2}$ and, thus,
go to infinity at $T\rightarrow T_0$.
The limiting temperature also corresponds to the softest point of the equation of state,
i.e., the speed of sound squared $c_s^2=dp/d\varepsilon$ goes to zero  as $(T_0-T)^{1/2}$.
Very similar thermodynamical behavior takes place in the Hagedorn model
for the special choice of a power, namely $m^{-4}$, in the pre-exponential factor of
the mass spectrum $\rho(m)$.
\end{abstract}

\maketitle

\section{Introduction}
\label{intr}

The van der Waals (VDW) equation of state (EoS) is a simple analytical model
of the pressure function $p$ for the system of  particles with both attractive
and repulsive interactions.
In the canonical ensemble (CE), where independent variables are temperature $T$,
volume $V$, and number of particles $N$, the VDW EoS has the most simple and
transparent form, (see, e.g., Refs.~\cite{greiner,LL}),
\eq{\label{eq:vdw}
p(T,n) ~=~ \frac{NT}{V-bN} ~-~ a\,\frac{N^2}{V^2}~ \equiv~\frac{n\,T}{1-bn}~-~a\,n^2~,
}
where $a>0$ and $b>0$ are the VDW parameters that describe attractive and
repulsive interactions, respectively, and $n\equiv N/V$ is the particle number
density.
The first term in the right-hand-side of Eq.~\eqref{eq:vdw} contains the excluded
volume correction
(for instance, $b=16 \pi r^3/3$ with $r$ being the particle hard-core radius),
the second term comes from the mean field description of the attractive interactions.

In order to apply the VDW EoS to systems with variable number of particles the
grand canonical ensemble (GCE) formulation is needed.
This was done for the VDW equation \eqref{eq:vdw} in our recent
paper \cite{VdW-GCE}.
The GCE formulation of EoS in the form $p=p(T,n)$,
including the VDW equation, can also be conveniently treated
within the thermodynamic mean-field approach~\cite{mf-1992,mf-1995,mf-2014}.
As the next step, we proposed in Ref.~\cite{VdW-NM} the generalization of the
VDW EoS that includes effects of the quantum statistics.
The nuclear matter was considered in \cite{VdW-NM} as the system of interacting
nucleons with Fermi statistics and the VDW EoS.
With this model the curve of the first order phase transition and its end point
(the so called critical point) were obtained.
In Ref.~\cite{VdW-kurtosis} the fluctuations in the vicinity of the critical point
were investigated.

In the present  paper we apply the VDW EoS with quantum statistics to
a description of the gas of interacting pions.
In this case the GCE formulation is really necessary:
the number of pions is not a conserved quantity and cannot be considered as an
independent variable of the CE.
Note, the EoS of interacting pions was studied during last decades
within different theoretical approaches (see, e.g.,~\cite{pion,pion1,pion2,pion3,pion4}
and references therein).

The paper is organized as follows.
In Sec.~\ref{sec-VDW} we present the VDW EoS that includes the effects of quantum statistics.
In Sec.~\ref{sec-pions} this formalism is applied to the pion gas.
The effects related to the limiting temperature are investigated in
Sec.~\ref{sec-lim} and a comparison with the Hagedorn model
is conducted in Sec.~\ref{sec-hag}.
A summary in Sec.~\ref{sec-sum} closes the article.

\section{VdW equation of state for quantum statistics}
\label{sec-VDW}

The pressure and the particle number density for the quantum VDW EoS in the GCE
are defined as the following \cite{VdW-NM}:
\begin{equation}
\label{n,p}
p(T,\mu)~=~p^{\rm id} (T, \mu^*) - a\,n^2(T,\mu)~,
\qquad
n(T,\mu)~=~\frac{n^{\rm id}(T,\mu^*)}{1 + b \, n^{\rm id}(T,\mu^*)}~,
\end{equation}
where
\begin{equation}
\label{mu^*}
\mu^*~ =~ \mu~ - ~b \, p(T,\mu) - a\,b\,n^2(T,\mu) + 2 \, a \, n(T,\mu)~,
\end{equation}
with $\mu$ being the chemical potential which regulates the particle number
density in the GCE.
Quantities $p^{\rm id}$ and $n^{\rm id}$ correspond to the quantum
ideal gas pressure and particle density, respectively:
\begin{eqnarray}
\label{p-id}
p^{\rm id}(T,\mu^*) &=& \frac g3 \int \frac{d^3k}{(2 \pi)^3}\,
\frac{\bvar k^2}{\sqrt{m^2+  \bvar k^2}} \,
\left[ \exp{\left(\frac{\sqrt{m^2+  \bvar k^2}-\mu^*}{T}\right)} + \eta\right]^{-1}~,
\\
n^{\rm id}(T,\mu^*) &=& g \int \frac{d^3k}{(2 \pi)^3}\,
\left[ \exp{\left(\frac{\sqrt{m^2+ \bvar k^2}-\mu^*}{T}\right)} + \eta\right]^{-1}~,
\label{n-id}
\end{eqnarray}
where $g$ is the degeneracy factor, $m$ is the particle mass, $\eta = +1$ for
Fermi statistics, $\eta = -1$ for Bose statistics.

The quantum formulation (\ref{n,p})--(\ref{n-id})
of the VDW EoS in the GCE has the following basic properties:
(i) it transforms to the ideal {\it quantum} gas
if both $a=0$ and $b=0$; (ii) it becomes equivalent
to the classical VDW EoS \eqref{eq:vdw}
in a region of thermodynamical parameters where quantum statistics can be neglected;
(iii) the entropy found from Eqs.~(\ref{n,p})--(\ref{n-id})
is a non-negative quantity and it goes to zero at $T\rightarrow 0$
in accordance with the third law of thermodynamics (see Ref.~\cite{VdW-NM} for details).
Note, in the Boltzmann approximation, i.e., $\eta=0$ in Eqs.~(\ref{p-id}) and (\ref{n-id}),
EOS obtained in the
GCE from (\ref{n,p})--(\ref{n-id}) as $p = p(T,n)$ becomes identical 
to the VDW equation of state given in Eq.~(\ref{eq:vdw}).

\section{Pion gas with the VDW EoS}\label{sec-pions}
In a case of the pion gas one should use the Bose statistics,  $\eta=-1$ in Eqs.(\ref{p-id})
and (\ref{n-id}). The pion system with zero value of total electric charge is considered,
thus, $\pi^+$, $\pi^-$, and $\pi^0$ all have zero chemical potentials, i.e.,
$\mu=0$ in Eq.~(\ref{mu^*}). As a result,
$p$ and $n$ in Eq.~(\ref{n,p}) become the functions of one variable, $T$,
and Eqs.~(\ref{mu^*})--(\ref{n-id}) take the form:
\begin{eqnarray}
\label{mu^*-bose}
\mu^*~ & = & - ~b \, p(T) - a\,b\,n^2(T) + 2 \, a \, n(T)~, \\
\label{p-id-bose}
p^{\rm id}(T,\mu^*) &=& \frac{1}{2\pi^2} \int_0^{\infty} k^2dk\,
\frac{ k^2}{\sqrt{m_{\pi}^2+  k^2}} \, \left[
\exp{\left(\frac{\sqrt{m_{\pi}^2+  k^2}-\mu^*}{T}\right)}\, -1\right]^{-1}~,
\\
n^{\rm id}(T,\mu^*) &=& \frac{3}{2\pi^2} \int_0^{\infty} k^2 dk\,
\left[ \exp{\left(\frac{\sqrt{m_{\pi}^2+ k^2}-\mu^*}{T}\right)} -1\right]^{-1}~,
\label{n-id-bose}
\end{eqnarray}
where we set $g=3$ for the pion degeneracy factor and
$m_{\pi}\cong 138~ {\rm MeV}$ for the pion mass.
In what follows we take a fixed value of $b\cong 0.45$~fm$^3$ that
corresponds to $r\cong0.3 ~{\rm fm}^{-3}$ for the pion hard-core radius
\cite{EV,HRG-VAG} \footnote{In the next section we discuss how the model
results depend on the numerical value of $b$, particularly at $b\rightarrow 0$.}.
The value of $a$ is considered as a free model parameter.
Note also that
the entropy density $s(T)$ and energy density $\varepsilon(T)$ can be calculated from
the function $p(T)$ as the following:
\eq{
\label{en}
s~=~\frac{dp}{dT}~,~~~~~~~\varepsilon(T)=T\frac{dp}{d T}
~-~p~.
}

We consider
$T\leq 160~{\rm MeV}$ as a region, where the pion gas may exist.
At higher temperatures the effects related to the deconfinement are expected to play a major role.
For every $T$, at $T\le 160$~MeV, we find $\mu^*=\mu^*(T)$  by solving numerically the
transcendental equation (\ref{mu^*-bose}), the functions $n(T)$ and $p(T)$ are then found
from Eq.~(\ref{n,p}).
If Eq.~(\ref{mu^*-bose}) has more then one solution for a given $T$,
only the solution with a larger value of  $p(T)$ should be considered as a stable one
according to the Gibbs criteria. For the VDW EoS in the GCE these
criteria were considered in Ref.~\cite{VdW-NM}

\begin{figure}
\includegraphics[width=0.49\textwidth]{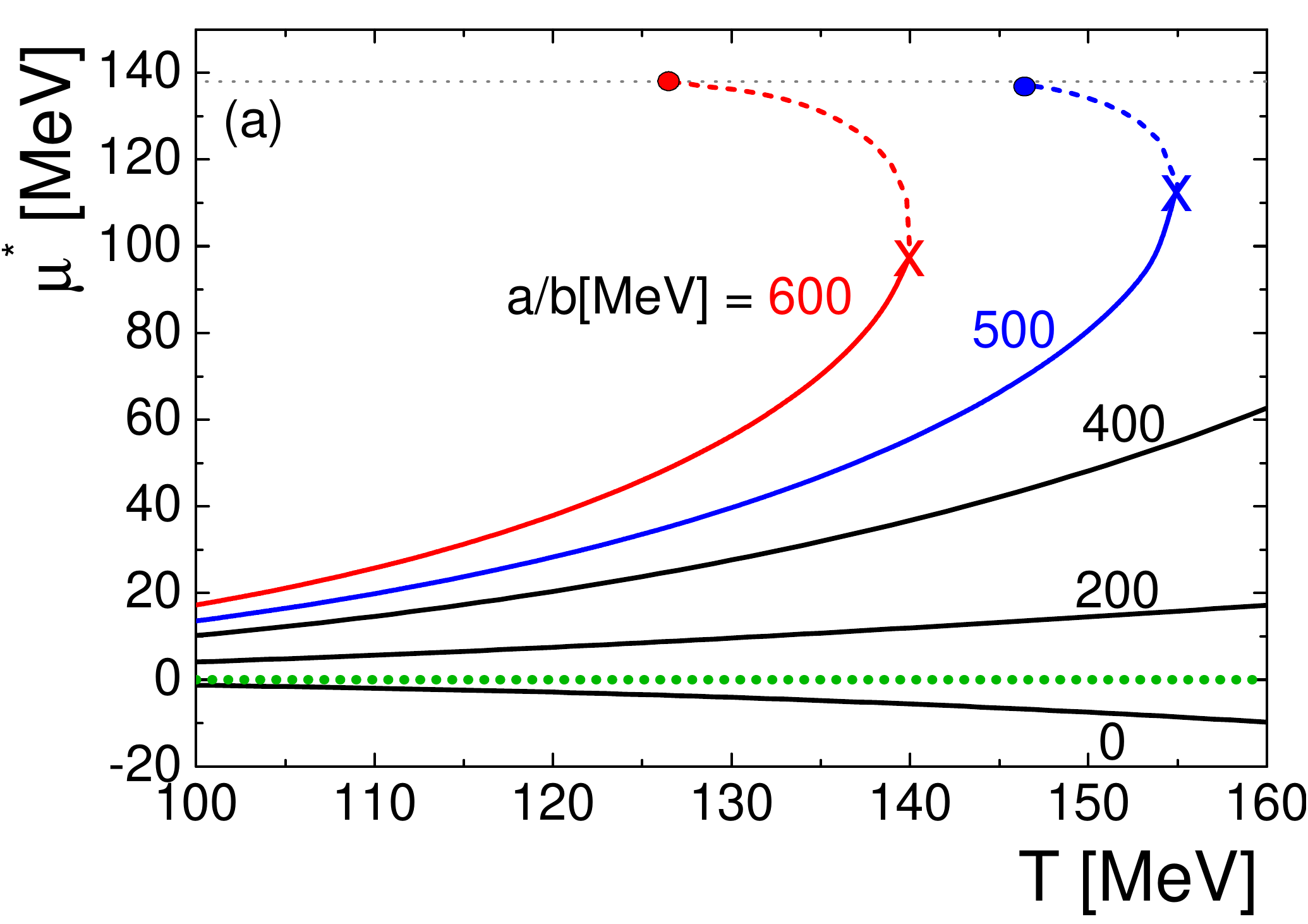}
\includegraphics[width=0.49\textwidth]{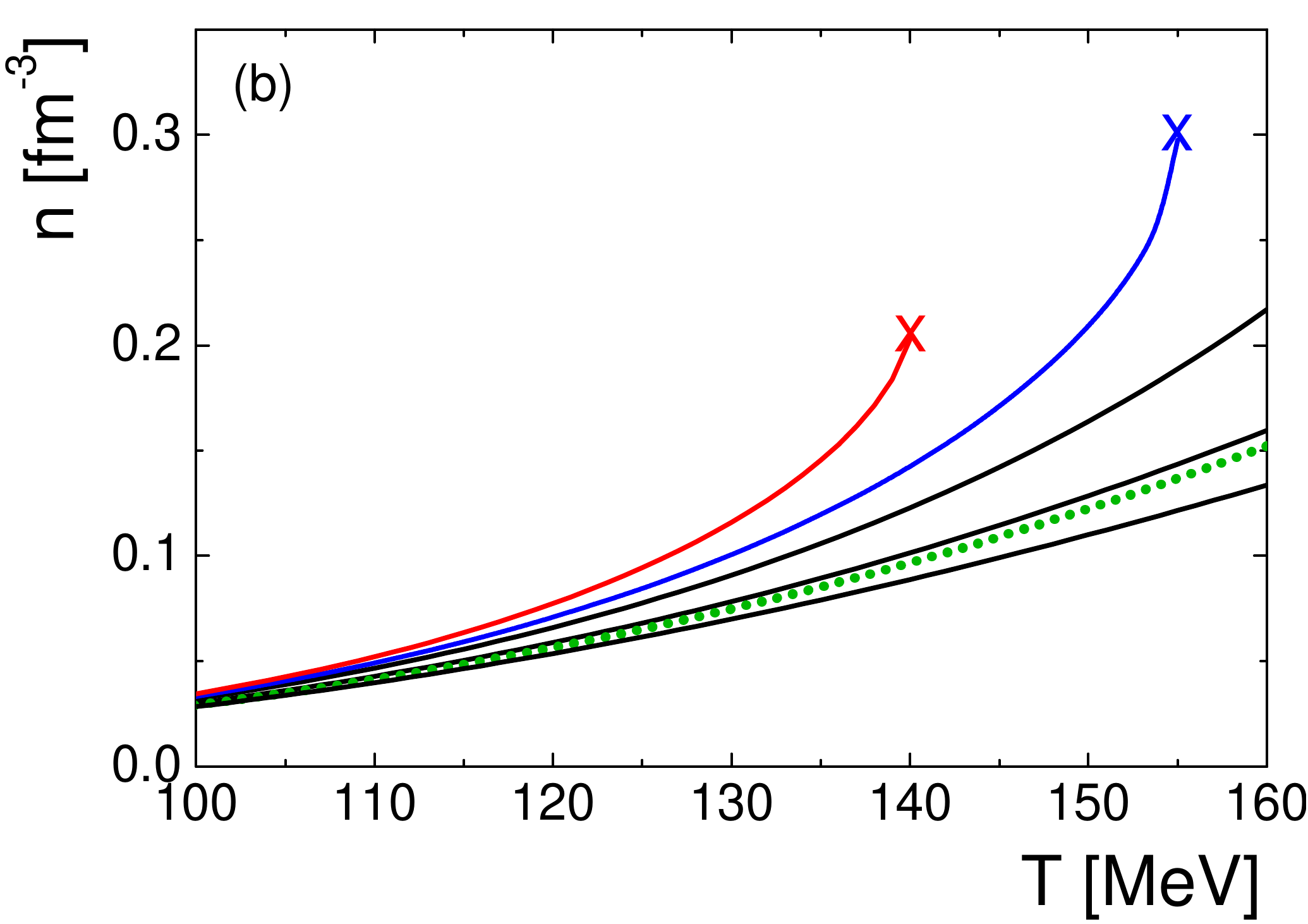}
\includegraphics[width=0.49\textwidth]{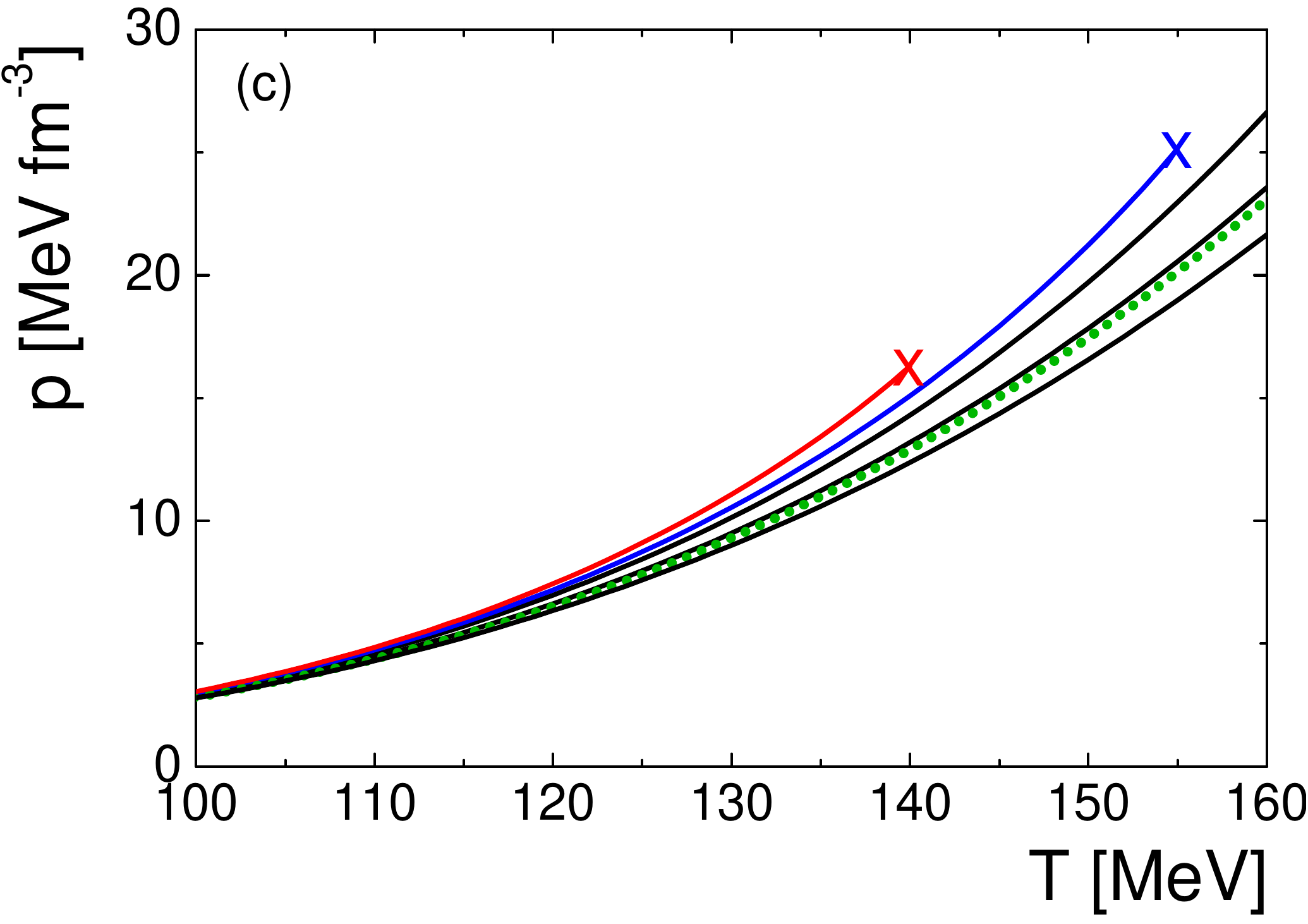}
\includegraphics[width=0.49\textwidth]{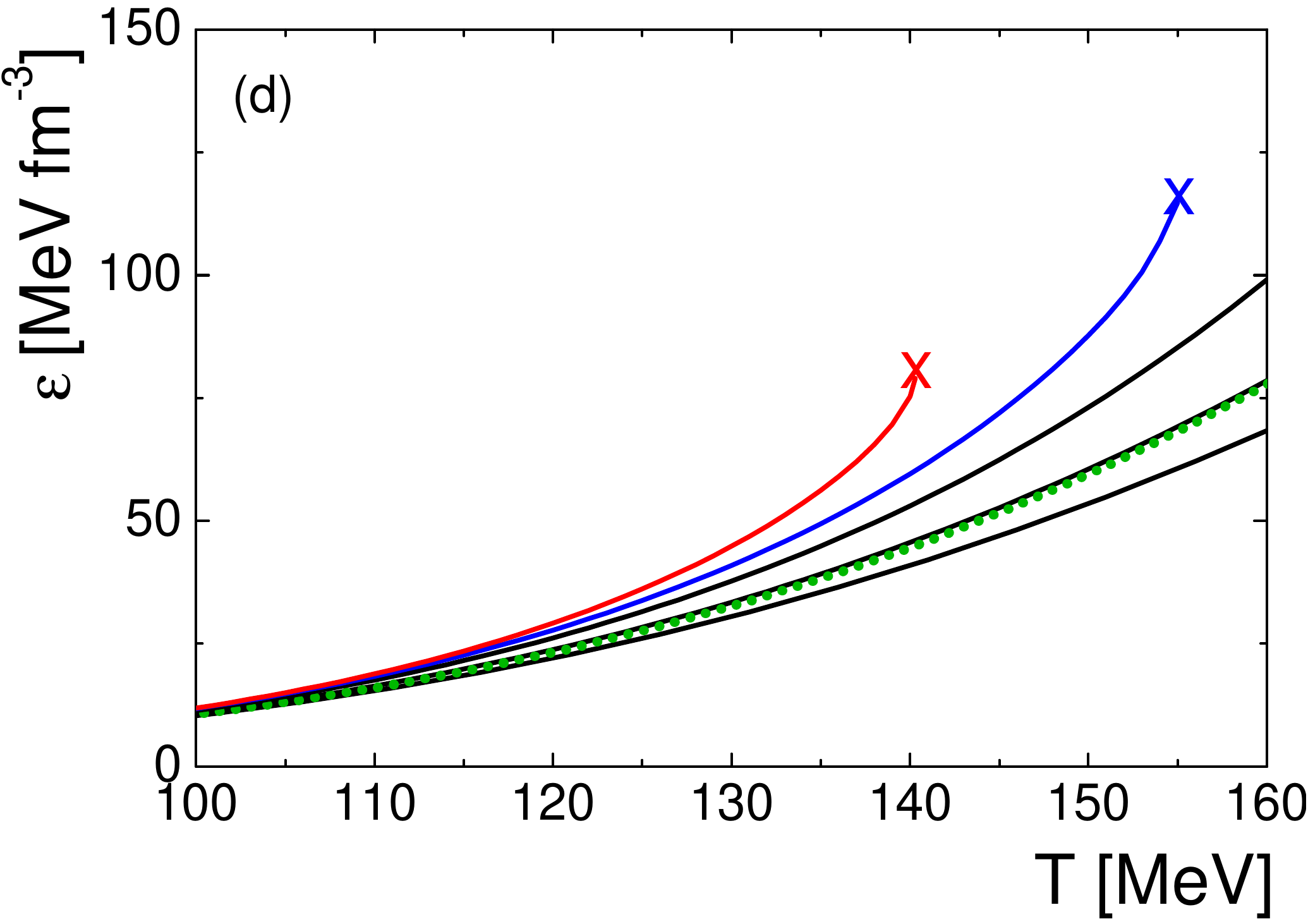}
\caption{The thermodynamical quantities of the pion gas with the VDW EoS
as functions of temperature: shifted chemical potential
$\mu^*$ (a), particle number density (b), pressure (c), and energy density (d).
The excluded volume parameter is fixed as $b=0.45$~fm$^3$. The solid lines
correspond to the solutions with $a/b$ = 0, 200, 400, 500, 600 MeV
(from bottom to top).
The dotted lines show the behavior of the ideal pion gas, i.e., for both $a=0$ and $b=0$.
The points at the limiting temperature are depicted by the cross symbols $\times$.
In the panel (a) the unstable solutions are depicted by dashed line segments
with endpoints at $T=T'$ depicted by the filled circles. }
\label{fig-p}
\end{figure}

The results for  $\mu^*$, $p$, $n$, and $\varepsilon$ as functions of $T$
are shown in Figs.~\ref{fig-p} (a)-(d) by solid lines for
different values of the parameter $a$.
Dotted lines in these figures correspond to the ideal pion gas (i.e., both $a=0$ and $b=0$).
In the region of $T\le 100$~MeV (not shown in Fig.~\ref{fig-p}) the interaction
effects for the considered $a$ and $b$ values are small, i.e., all lines are very close
to those of the ideal gas. This is due to the fact that the VDW interactions depend on the particle number density,
and they become negligible at $n\rightarrow 0$.

The repulsive interactions
suppress the VDW thermodynamic functions
$n(T)$, $p(T)$, and $\varepsilon(T)$ in comparison to those of the
ideal pion gas. This is clearly seen for lines with $a=0$ in Figs.~\ref{fig-p} (b)-(d).
The VDW repulsion also introduces the upper limit for the particle
number density, namely, $n\le 1/b \cong 2.21~{\rm fm}^{-3}$.
On the other hand, the presence of the attractive interactions enhances $n(T)$, $p(T)$,
and $\varepsilon(T)$ in comparison to the ideal pion gas.
One observes that
the VDW EoS with $b=0.45$~fm$^3$ and
$a/b = 200$~MeV appears to be close to that
of the ideal gas, i.e., the actions of repulsive and attractive
interactions almost cancel each other out in all thermodynamical functions.

We note that the VDW excluded volume correction
is consistent with the virial expansion for hard spheres only at low enough densities.
At large values of the so-called packing fraction $\eta = (b n) / 4$
(a fraction of the total volume occupied by the particles of finite size), namely $\eta \gtrsim 0.1$,
the VDW equation deviates significantly from the equation of state for hard spheres.
The highest value of the pion number density obtained in the present work
is $n_{\rm max} \simeq 0.30$~fm$^{-3}$, as seen in Fig.~\ref{fig-p}.
Therefore, the maximum value of the packing fraction is around $\eta_{\rm max} = 0.034$.
At such values of $\eta$ the VDW equation is still fully consistent with the virial
expansion for the hard spheres (see, e.g., Fig.~1 in Ref.~\cite{mf-2014}) and thus can be used.

\section{Limiting Temperature}
\label{sec-lim}

At large values of the parameter $a$ a new phenomenon takes place: a limiting
temperature $T_0$ emerges, i.e., Eq.~(\ref{mu^*-bose}) has no stable solutions at $T>T_0$
\footnote{Note that only temperatures $T<160$ MeV are considered.}.
At the same time in the temperature interval $T'\le T\le T_0$ the unstable
solutions for the VDW thermodynamical functions appear.
Here the temperature $T'$ is determined as a starting point of
the Bose-Einstein condensation, i.e., $\mu^*(T')=m_\pi$.
The values of the thermodynamical functions at the limiting temperature $T_0$
are depicted by the cross symbols $\times$  in  Figs.~\ref{fig-p} (a)-(d), while the
values at $T=T'$ are depicted in Fig.~\ref{fig-p} (a) by circles, and
the unstable solutions are shown by the dashed segments of lines.
The stable solutions shown in all figures by solid lines correspond to the Gibbs
criteria, i.e., they have pressure larger than that of the unstable
solutions at equal temperatures.
The unstable solutions are shown only for a shifted
chemical potential $\mu^*(T)$, in the panel (a) of Fig.~\ref{fig-p}.

\begin{figure}
\includegraphics[width=0.49\textwidth]{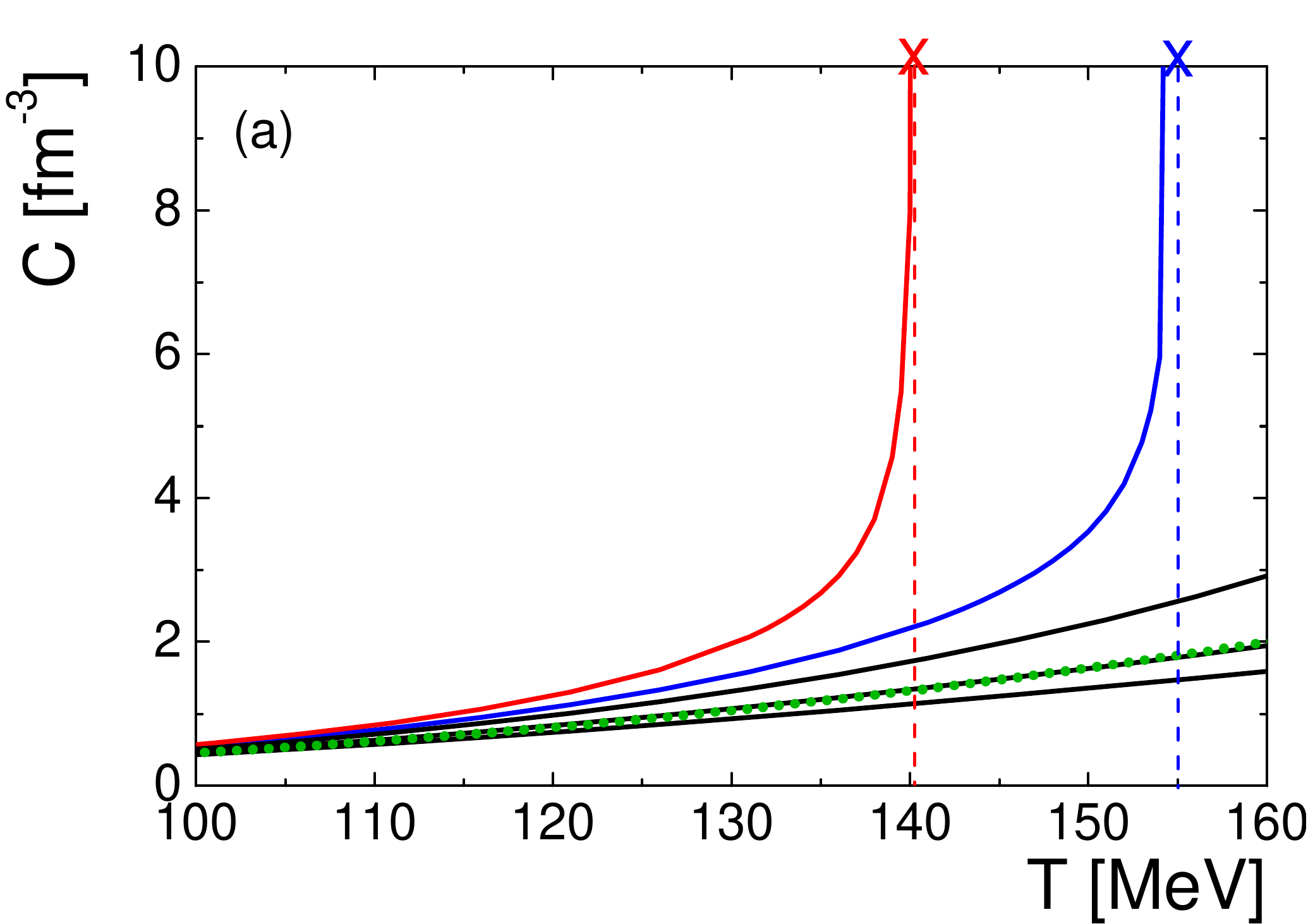}
\includegraphics[width=0.49\textwidth]{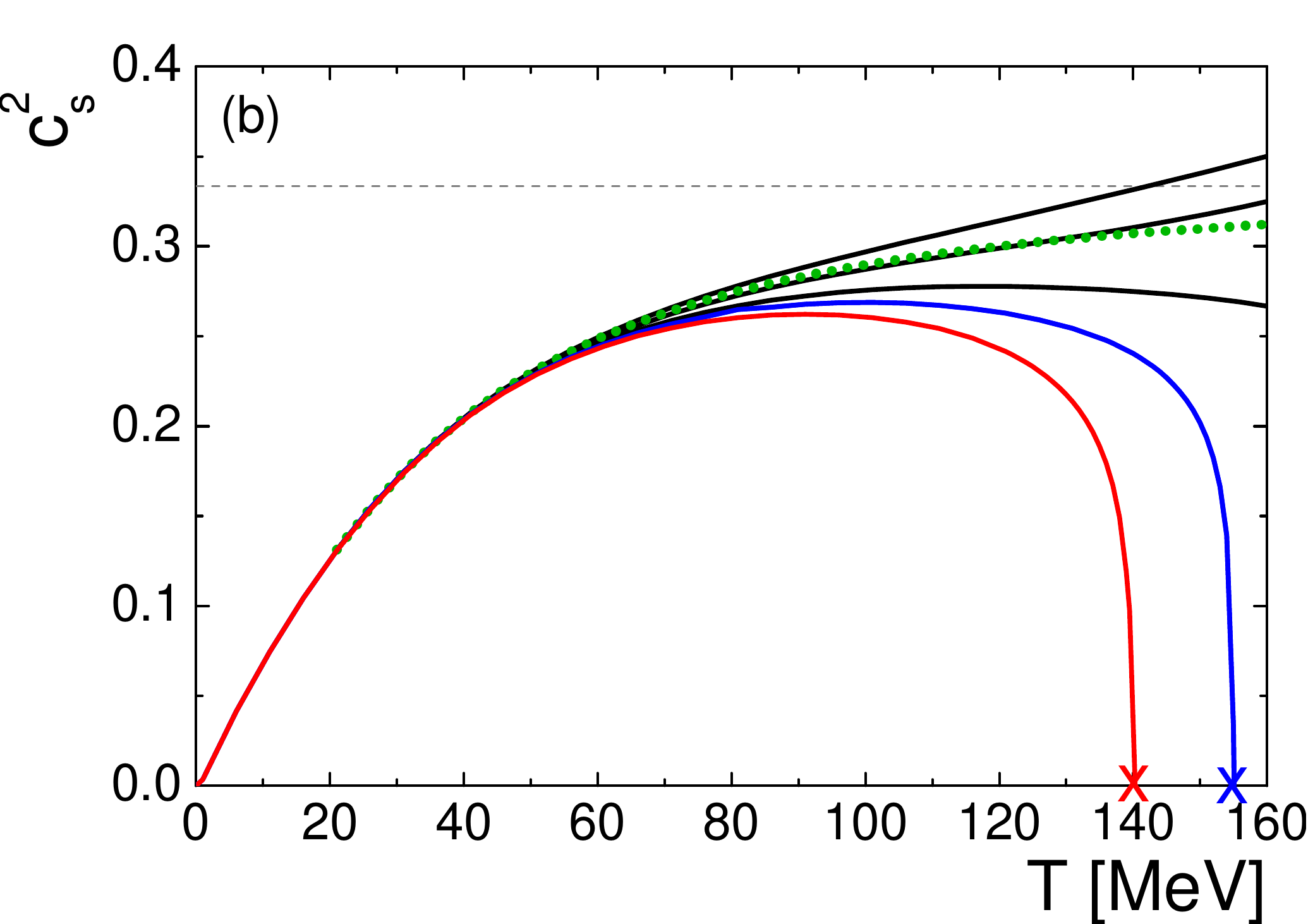}
\includegraphics[width=0.49\textwidth]{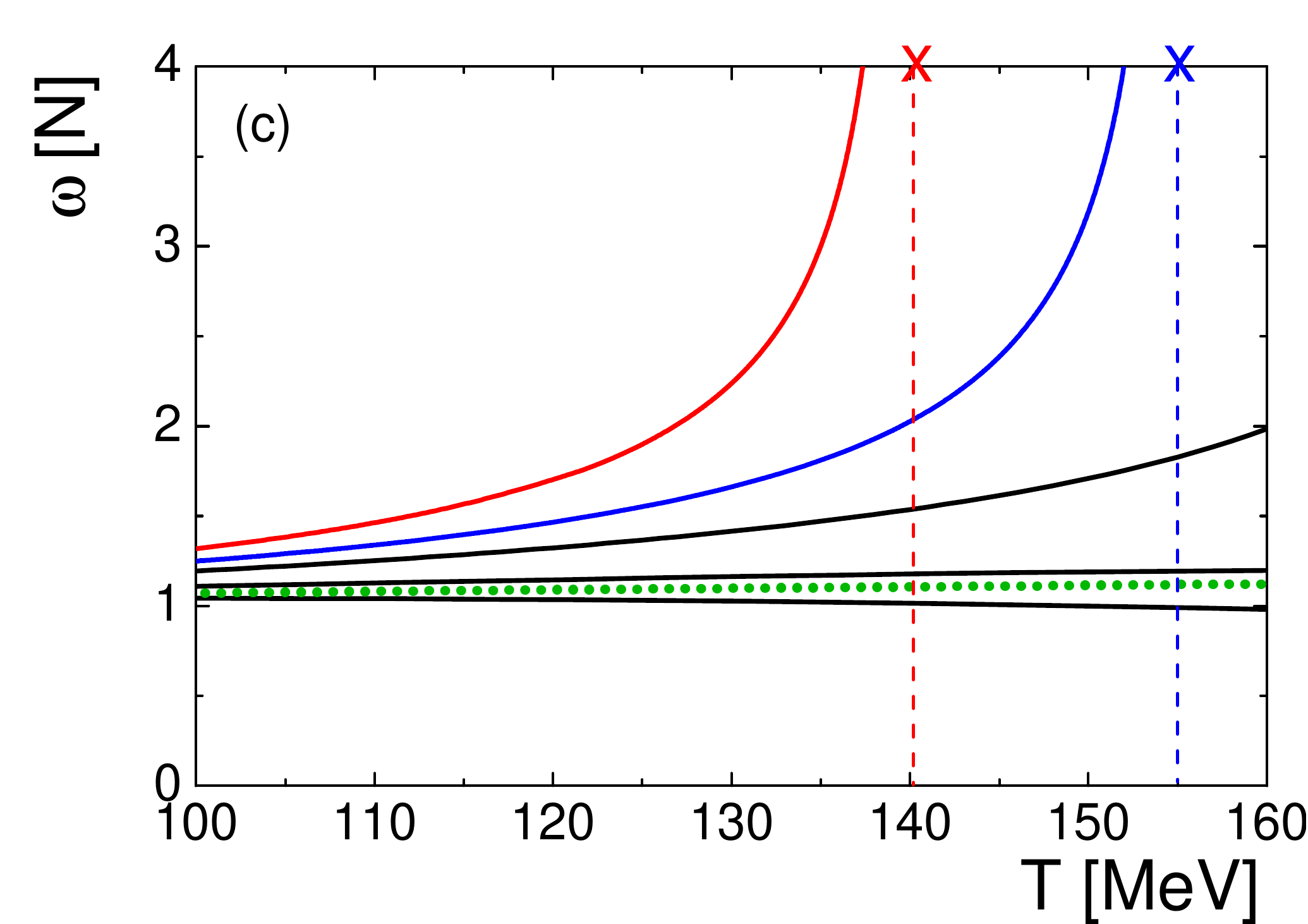}
\includegraphics[width=0.49\textwidth]{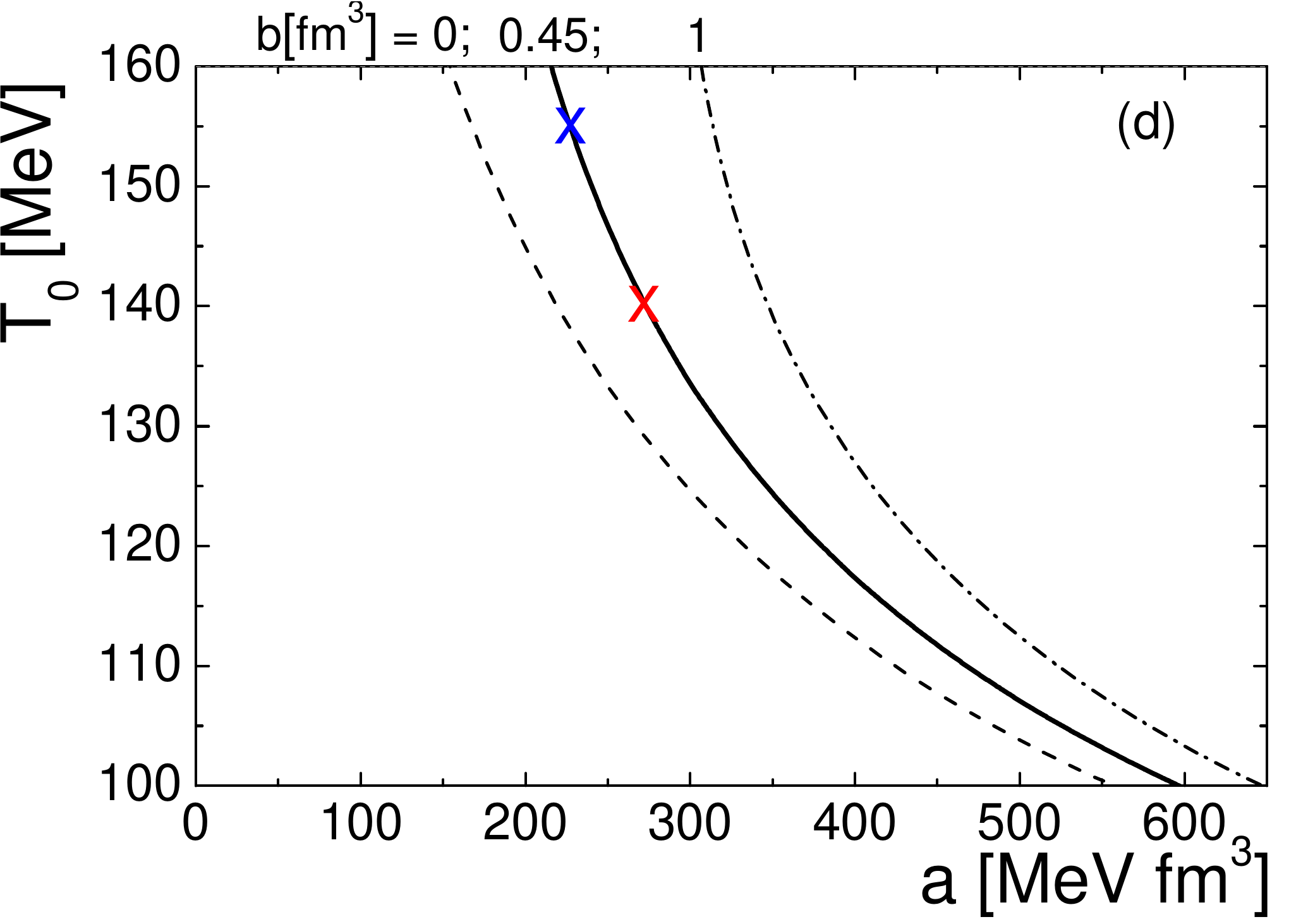}
\caption{
(a) The heat capacity.
(b) The speed of sound squared (the dotted horizontal line corresponds to $c_s^2=1/3$, obtained for the ideal gas of massless particles).
(c) The  scaled variance $\omega[N]$ (\ref{w}).
The solid lines correspond to solutions with
$b=0.45$~fm$^3$ and $a/b=$ 0, 200, 400, 500, 600~MeV from bottom to top in (a) and (c),
and from top to bottom in (b).
(d) A dependence of the limiting temperature $T_0$ on the parameter $a$ for
$b=0$ (dashed  line), $b=0.45$~fm$^3$ (solid line), and $b=1$~fm$^3$ (dashed-dotted line).
}
\label{fig-w}
\end{figure}

To elucidate a physical origin of the limiting temperature $T_0$ for the pion gas with
the VDW EoS we calculate the specific heat, $C=d\varepsilon/dT$,
the speed of sound squared, $c_s^2=dp/d\varepsilon$,
and the scaled variance of particle number fluctuations,
\begin{equation}
\label{w}
\omega[N]\, =\, \frac{\langle N^2 \rangle - \langle N \rangle^2}
{\langle N \rangle}\, =\, \omega_{{\rm id}}(T,\mu^*)\left[\frac{1}{(1-b n)^2}
- \frac{2 a n(T,\mu^*)}{T}\, \omega_{\rm id}(T,\mu^*)\right]^{-1} \,,
\end{equation}
where
\begin{equation}
\label{wid}
\omega_{\rm id}(T,\mu^*)\, =\, \frac{1}{n^{\rm id}(T,\mu^*)}~\frac{3}{2\pi^2}
\int_0^{\infty} k^2 dk\,
\left[\exp{\left(\frac{\sqrt{m_\pi^2+k^2}-\mu^*}{T}\right)}~-~1\right]^{-2}~,
\end{equation}
is the scaled variance of the particle number fluctuations in the ideal pion gas with the chemical potential $\mu^*$.
Equations~(\ref{w}) and (\ref{wid}) were obtained for the quantum VDW EoS in
Ref.~\cite{VdW-kurtosis}.
As seen from Figs.~\ref{fig-w} (a)-(c),
$c_s^2\rightarrow 0$, while both $C\rightarrow \infty$
and $\omega[N]\rightarrow\infty$ at $T\rightarrow T_{0}$.
The straightforward calculations reveal the power-law behavior at
$T\rightarrow T_0$: both $C$ and $\omega[N]$ are proportional to $(T_0-T)^{-1/2}$,
and $c_s^2 \propto (T_0-T)^{1/2}$.
Fluctuations of the total energy $E$ behave very similar to fluctuations of the
number of pions $N$.
The limiting temperature corresponds, therefore, to the softest point of the EoS
(the speed of sound equals zero) and to infinitely large values of the heat
capacity and fluctuations.
In Fig.~\ref{fig-w} (d) a dependence of $T_0$ on parameter $a$ is shown
for different values of repulsive parameter $b$:
$b=0$, $b=0.45$~fm$^3$ and $b=1$~fm$^3$.
It is seen from Fig.~\ref{fig-w} (d) that, at each value of VDW
repulsion parameter $b$, there is a minimal value of VDW attraction parameter $a=a_b$ such that
the limiting temperature
$T = T_0 \leq 160$~MeV appears at all $a\ge a_b$, and $T_0$ decreases with increasing $a$.
At $b=0.45$~fm$^3$, one finds that $T_0\cong 140$~MeV and 155~MeV at $a\cong 270$~MeV~fm$^3$
and 225~MeV~fm$^3$, respectively. For the smaller values of $b$,
the same values of the limiting temperature, $T_0\cong 140$~MeV and 155~MeV, are reached
at smaller values of $a$. Particularly, for $b=0$,  the values of
$T_0\cong 140$~MeV and 155~MeV are obtained at $a\cong 219$~MeV~fm$^3$
and 168~MeV~fm$^3$, respectively. Note that additional information is needed
to have further restrictions on the parameters $a$ and $b$. From the
present analysis the most interesting physical behavior with $T_0=150\pm 10$~MeV
takes place in the range
of $b$ from 0 to 0.45~fm$^3$ and of $a$ from 150~MeV~fm$^3$ to 300~MeV~fm$^3$.

The VDW parameters $a$ and $b$ were  obtained in Ref.~\cite{VdW-NM} for a system of nucleons.
Based on the ground state properties of the nuclear matter it has been estimated that
for nucleons $b \simeq 3.42$~fm$^3$ and $a \simeq 329$~MeV~fm$^3$.
Therefore, the numerical values of both  $b$ and $a$ VDW parameters 
in the pion gas considered in the present
study are smaller than those in the system of interacting nucleons.

In principle, the information about the VDW parameters $a$ and $b$ could be inferred
from the $\pi\pi$-scattering data.
Note that second cluster integral for the equilibrium pion gas
was calculated in Ref.~\cite{KG}. The phase shifts in $\pi\pi$
scattering were described by assuming that both
the hard-core repulsion and resonance attraction are equally important 
and should be taken into account simultaneously.
From the specific calculation in Ref.~\cite{KG} it could be inferred
that parameter $a$ may take values up to few hundred
MeV~fm$^3$, and could in general be temperature dependent.
However, there is no reason to expect that the higher terms in
the cluster expansion are negligible.
The attractive parts of higher cluster terms can be presented 
as contributions of multi-pion resonances.
The properties of the resonances 
(masses and widths), and even their existence
themselves, are not very well known. Therefore, an accurate estimation 
of the numerical value of parameter $a$ in the VDW system of pions 
looks rather problematic, and it is treated as purely phenomenological 
free parameter in the present study.

\section{Comparison with the Hagedorn model}
\label{sec-hag}

A concept of the limiting temperature for the hadron gas was introduced in
physics by Hagedorn \cite{hag,hag-1}.
His limiting temperature $T_0 =160\pm 10$~MeV emerged due to the exponentially
increasing mass spectrum $\rho(m)$ for the hadron excited states at large $m$:
\eq{\label{rho}
\rho(m)~\cong ~c\,m^{-\alpha}\,\exp\left(\frac{m}{T_0}\right)\,\theta(m-M_0)~,
}
where $c$, $M_0$, $T_0$, and  $\alpha$ are the model parameters.
These excited states called fireballs were considered as a point-like
non-interacting particles.
A presentation of particle attractive interactions by the resonance states
was first suggested in Ref.~\cite{BU} in statistical physics
and then applied in Ref.~\cite{Bel} to hadron production.
This idea was then further  extended within the $S$-matrix formulation of
statistical mechanics developed in Ref.~\cite{DMB}.

The spectrum $\rho(m)$ can be found from the statistical bootstrap equation \cite{fraut}.
This equation requires the spectrum of low-lying  hadron states as an input.
In a simplest version of the model, this input is reduced to a single lightest
hadron -- the pion, and the continuous mass spectrum $\rho(m)$ starting from $M_0> 2m_{\pi}$.
The pressure of the Hagedorn model
is then defined as
\eq{\label{Hp}
p_{\rm H}(T)~& =~ \int \frac{d^3k}{(2\pi)^3}\,\frac{\bvar k^2}{\sqrt{\bvar k^2+m_\pi^2}}\,
\left[\exp{ \left(\frac{\sqrt{\bvar k^2 + m_\pi^2}}{T}\right)}~-~1\right]^{-1}
\nonumber \\
&+~\frac{T^2}{2\pi^2}\int_{M_0}^{\infty}dm\,\rho(m)\,m^2K_2(m/T)~,
}
where $K_2$ is the modified Bessel function.
The first term in the right hand side of Eq.~(\ref{Hp}) corresponds to the pressure
of the ideal Bose gas of pions and the second one to the contribution
of all excited states (\ref{rho}) taken in the Boltzmann approximation.
Other thermodynamical functions of the Hagedorn model are then given by the
thermodynamical relations (\ref{en}).

\begin{figure}
\includegraphics[width=0.49\textwidth]{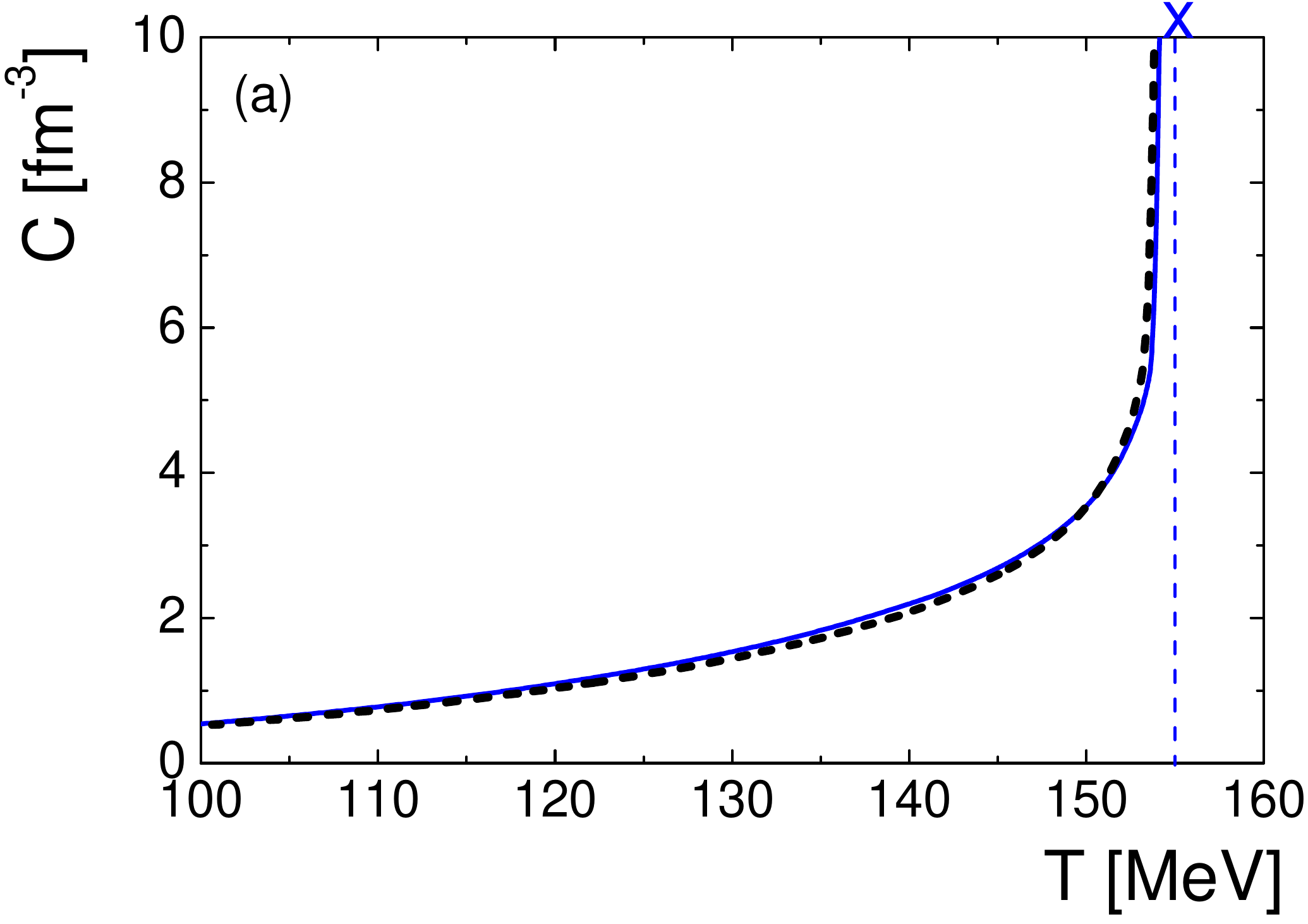}
\includegraphics[width=0.49\textwidth]{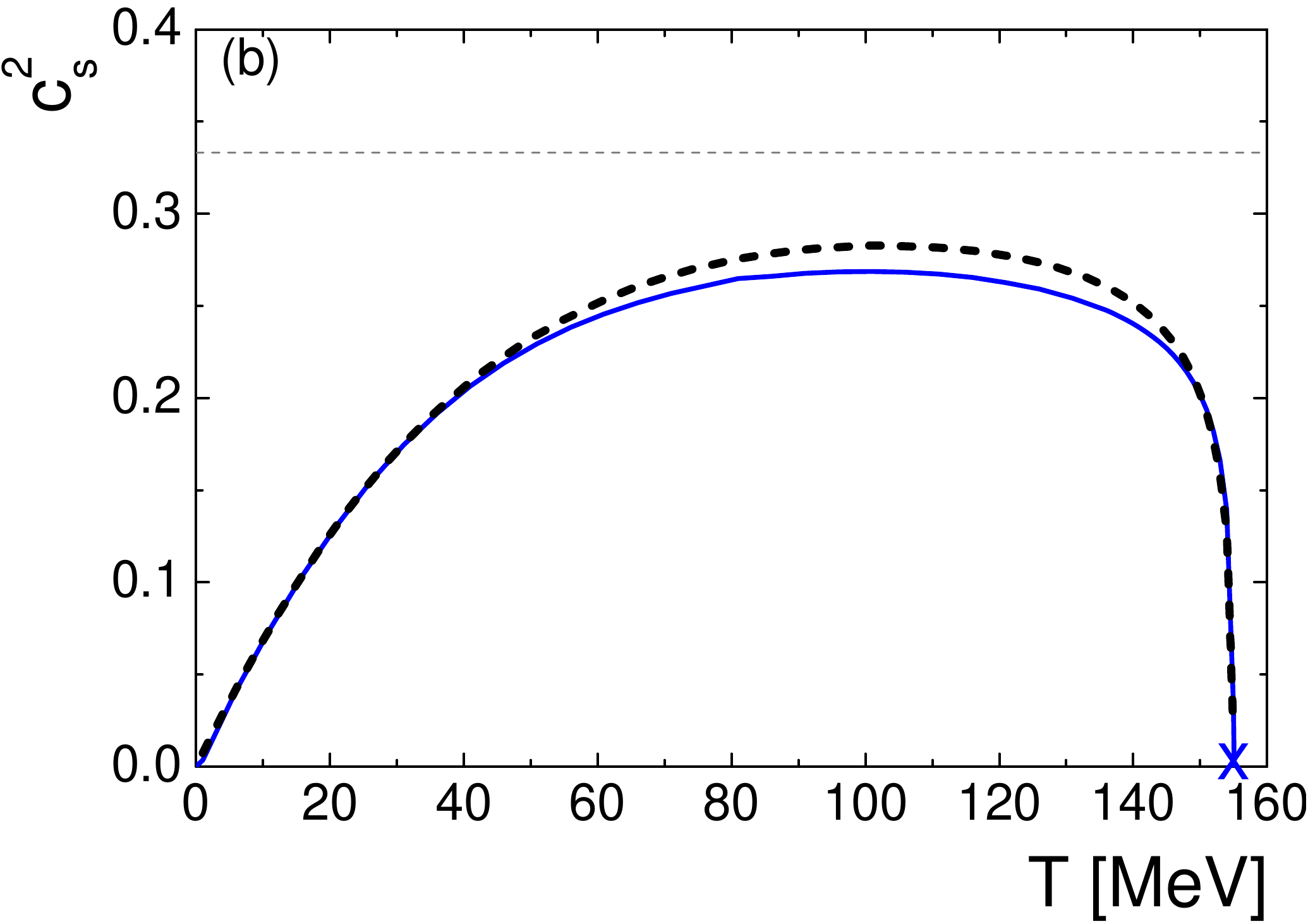}
\caption{
The heat capacity (a) and the speed of sound (b). The solid  lines
correspond the VDW EoS with
$b=0.45$~fm$^3$ and $a/b=500$~MeV. The dotted lines correspond to the Hagedorn
model (\ref{rho}) and (\ref{Hp}) with $T_0=155$~MeV, $\alpha=4$, $c=3.57$~fm$^{-3}$,
and $M_0=507$~MeV.
}
\label{fig-hag}
\end{figure}

From Eqs.~(\ref{rho}) and (\ref{Hp}) it follows that the thermodynamical functions
do not exist at $T>T_0$ because of a divergence of the integral with respect to $m$
in Eq.~(\ref{Hp}).
The singular part of the integral contributes to the pressure
as $p_H\propto (T_0-T)^{\alpha -5/2}$
and to the energy density as $\varepsilon_H \propto (T_0-T)^{\alpha-7/2}$.
An explicit solution of the statistical
bootstrap equation \cite{fraut} gives $\alpha =3$ \cite{yellin}.
Thus, $\varepsilon_H \propto (T_0-T)^{-1/2}$ and $\varepsilon_H\rightarrow \infty$
at $T\rightarrow T_0$. On the other hand,
for $\alpha > 7/2$ all thermodynamical functions in the Hagedorn model
remain finite in a vicinity of
the limiting temperature $T=T_0$.
Even more, there is a remarkable correspondence
between the pion gas with
the VDW EoS and the Hagedorn model with $7/2<\alpha <9/2$,
i.e., the behavior
of  $p_H$ and $\varepsilon_H$ is very similar to that
of $p$ and $\varepsilon$ of the VDW EoS shown in the panels (b) and (c)
of Fig.~\ref{fig-p} for $a/b=500$~MeV and 600~MeV.
In the vicinity of the
limiting temperature $T_0$
the heat capacity and the speed of sound squared behave in Hagedorn model
as $C\propto (T_0-T)^{\alpha-9/2}$ and $c_s^2\propto (T_0-T)^{9/2-\alpha}$, respectively.
Therefore, for $\alpha=4$ the behavior of $C$ and $c_s^2$ in the Hagedorn model
at $T\rightarrow T_0$ is the same as in the pion gas with the VDW EoS, if the $T_0$
values in both models are set to be equal.  In Fig.~\ref{fig-hag} these two quantities
are compared for the VDW model with $b=0.45$~fm$^3$, $a/b=500$~MeV and the Hagedorn
model with $T_0=155$~MeV and $\alpha=4$.

%

\section{Summary}
\label{sec-sum}

In summary, the interacting pion system has been studied using the quantum
version of the VDW EoS within the GCE.
The role of repulsive and attractive interactions described by the VDW parameters
$b$ and $a$, respectively, was considered.
It is found that for $b\cong 0.45$~fm$^3$ and $a/b\cong 200$~MeV the VDW EoS for
pions appears to be close to that of the ideal pion gas,
i.e., repulsive and attractive interactions are approximately canceled out.
At each value of $b$ there is a minimal value of $a=a_b$ such that the limiting temperature
$T=T_0$ appears at all $a\ge a_b$, and $T_0$ decreases with increasing $a$.
When $T\rightarrow T_0$ several remarkable  effects happen: the speed of sound
approaches zero as $(T_0-T)^{1/2}$, while the heat capacity as well as the
scaled variance of fluctuations of the particle multiplicity
go to infinity as $(T_0-T)^{-1/2}$.

The presented VDW model for pions is compared to the Hagedorn model,
where the phenomenon of limiting temperature is also present.
Even though the realization of the attractive mechanism between particles
in the Hagedorn model is different from that in the VDW model,
we found that a very similar thermodynamical behavior emerges in both models.
It takes place for the special choice of a power, namely $m^{-4}$, in the
pre-exponential factor of the mass spectrum $\rho(m)$ of the Hagedorn model.

A presence of the limiting temperature $T_0$ 
in the VDW pion gas is definitely a signal of the restricted validity of this model.
Similar to the Hagedorn model, the limiting temperature of the VDW pion system
should be transformed to the temperature
of deconfinement transition when the fundamental quark-gluon degrees of freedom 
are introduced. 
This consideration is, however, outside of 
the scope of the present paper.

\begin{acknowledgments}
We are thankful to M. Ga\'zdzicki for fruitful comments and discussions.
This work was supported
by the Program of Fundamental Research
of the Department of Physics and Astronomy of National Academy of Sciences of Ukraine,
and by HIC for FAIR within the LOEWE program of the State of Hesse.
\end{acknowledgments}


\end{document}